\documentclass[12pt]{article}
\usepackage{amscd,amssymb,amsmath,latexsym,enumerate}
\usepackage[mathscr]{euscript}
\usepackage{epsfig}
\usepackage{fancybox}
\usepackage{verbatim}
\usepackage{stmaryrd}

\usepackage[hidelinks]{hyperref}
\hypersetup{ 
       colorlinks=true,
       linkcolor=black, 
       citecolor=black,
       filecolor=black,
       menucolor=black,
       urlcolor=black,
       breaklinks=true
}

\usepackage[small,nohug]{diagrams}
\diagramstyle[labelstyle=\scriptstyle]

\usepackage{color}

\title{Dimensional reduction and scattering formulation \\ for even topological invariants}

\author{Hermann Schulz-Baldes, Daniele Toniolo
\\
\\
{\small Department Mathematik, Friedrich-Alexander-Universit\"at Erlangen-N\"urnberg, Germany}
}

\date{ }

\newtheorem{theo}{Theorem}
\newtheorem{defini}{Definition}
\newtheorem{proposi}{Proposition}
\newtheorem{lemma}{Lemma}
\newtheorem{coro}{Corollary}

\newcommand{\BM}{{\mathbb B}}
\newcommand{\CM}{{\mathbb C}}
\newcommand{\NM}{{\mathbb N}}
\newcommand{\RM}{{\mathbb R}}
\newcommand{\SM}{{\mathbb S}}
\newcommand{\TM}{{\mathbb T}}
\newcommand{\ZM}{{\mathbb Z}}
\newcommand{\PM}{{\mathbb P}}

\newcommand{\HM}{{\mathbb H}}
\newcommand{\DM}{{\mathbb D}}

\newcommand{\UM}{{\mathbb U}}

\newcommand{\Aa}{{\cal A}}
\newcommand{\Ee}{{\cal E}}

\newcommand{\Bb}{{\cal B}}

\newcommand{\Ff}{{\cal F}}
\newcommand{\Gg}{{\cal G}}

\newcommand{\Oo}{{\cal O}}
\newcommand{\Tt}{{\cal T}}

\newcommand{\Nn}{{\cal N}}
\newcommand{\Mm}{{\cal M}}
\newcommand{\Cc}{{\cal C}}
\newcommand{\Jj}{{\cal J}}

\newcommand{\Kk}{{\cal K}}
\newcommand{\Hh}{{\cal H}}

\newcommand{\one}{{\bf 1}}

\newcommand{\Tr}{\mbox{\rm Tr}}

\newcommand{\Ch}{{\rm Ch}}

\newcommand{\Exp}{{\rm Exp}} 
\newcommand{\Ker}{{\rm Ker}} 
\newcommand{\Ran}{{\rm Ran}}

\newcommand{\diag}{{\rm diag}}

\newcommand{\HScat}{H_{\mbox{\rm\tiny Scat}}}
\newcommand{\HCoup}{H_{\mbox{\rm\tiny Coup}}}

\newcommand{\HWire}{H_{\mbox{\rm\tiny Wire}}}
\newcommand{\HWireH}{\widehat{H}_{\mbox{\rm\tiny Wire}}}

\newcommand{\HIns}{H_{\mbox{\rm\tiny Ins}}}
\newcommand{\HInsH}{\widehat{H}_{\mbox{\rm\tiny Ins}}}

\begin{document}

\maketitle

\vspace{-1.0cm}

\begin{abstract}
Strong invariants of even-dimensional topological insulators of independent Fermions are expressed in terms of an invertible operator on the Hilbert space over the boundary. It is given by the Cayley transform of the boundary restriction of the half-space resolvent. This dimensional reduction is routed in new representation for the $K$-theoretic exponential map. It allows to express the invariants via the reflection matrix at the Fermi energy, for the scattering set-up of a wire coupled to the half-space insulator.   
\end{abstract}

\vspace{-.5cm}

\section{Introduction}

Dimensional reduction has been a source of inspiration in the field of topological insulators in quite diverse contexts. It plays a prominent role in the field theoretic framework \cite{QHZ} and it is invoked for the explanation of the periodic table \cite{RSFL,SCR}. Clearly also the bulk-boundary correspondence \cite{EG,KRS,PS} can be seen as a dimensional reduction as it allows to connect $d$-dimensional topological bulk invariants to boundary invariants. However, in its $K$-theoretic formulation \cite{KRS,PS} these boundary invariants are naturally expressed in terms of operators decaying away from the lower dimensional boundary, rather than being strictly supported by the boundary. This work establishes this last gap in a rigorous formulation of dimensional reduction for tight-binding Hamiltonians, namely it shows how the boundary invariants can be calculated from suitable operators defined on the Hilbert space over the lower dimensional boundary. Let us state right away the main result of the paper, even though the definitions of the Chern numbers as well as the precise assumptions and notations are only given in Section~\ref{sec-DimRedu}.


\begin{theo}
\label{theo-DimReduction} Let $d$ be even and $H=H_0+\lambda H_1$ be a covariant family of nearest neighbor one-particle Hamiltonians on $\ell^2(\ZM^d,\CM^L)$ which are a (possibly random) perturbation $\lambda H_1$ of an operator $H_0$ that is periodic in the first $d-1$ directions of $\ZM^d$. Suppose that $\mu\in\RM$ lies in a gap of the spectrum of $H$. Then the strong invariant $\Ch_d(P)$ of the Fermi projection $P=\chi(H\leq\mu)$ is given by
\begin{equation}
\label{eq-main}
\Ch_d(P)
\;=\;
-\,\Ch_{d-1}\big(
(\widehat{G}^{\mu+\imath\delta}\,-\,\imath\,\one)(\widehat{G}^{\mu+\imath\delta}\,+\,\imath\,\one)^{-1}\big)
\;,
\end{equation}
where $\delta>0$ is sufficiently small and $\widehat{G}^z=\Pi_1 (\widehat{H}-z)^{-1}\Pi_1^*$ the restriction to the boundary Hilbert space $\ell^2(\ZM^{d-1}\times\{1\},\CM^L)$ of the resolvent of the half-space restriction $\widehat{H}$ of $H $ to $\ell^2(\ZM^{d-1}\times\NM,\CM^L)$. 
\end{theo}

The proof of Theorem~\ref{theo-DimReduction} will be given in Section~\ref{sec-DimRedu}. It also provides a new proof for the two-dimensional case which was essentially already obtained in \cite{ASV} based on eigenvalue counting by intersection theory. The operator 
$$
\widehat{V}^z
\;=\;
(\widehat{G}^{z}\,-\,\imath\,\one)(\widehat{G}^{z}\,+\,\imath\,\one)^{-1}
$$
appearing on the r.h.s. of \eqref{eq-main} acts on a Hilbert space $\ell^2(\ZM^{d-1}\times\{1\},\CM^L)$ over the boundary (hyper-)surface and it only invokes the half-space Green matrix of the insulator. It is shown in Theorem~\ref{theo-ExpImage} below that the $K_1$-class of $\widehat{V}^z$ is the image of the Fermi projection $P$ under the $K$-theoretic exponential map. One of the delicate technical issues below is the limiting behavior of $\widehat{V}^z$ as $\Im m(z)\to 0$. Our somewhat unsatisfactory treatment in Section~\ref{sec-LimitCayley} below leads to the technical hypothesis stated in Theorem~\ref{theo-DimReduction}. Let us note that if $\widehat{H}=\widehat{H}_0$ is translation invariant along this surface, then the Green matrix $\widehat{G}^{z}$ can readily be linked to the well-known Weyl-Titchmarch function and it is shown that the boundary limit $\Im m(z)\to 0$ exists and is a unitary operator. If furthermore $H$ is also periodic in the direction perpendicular to the boundary, then $\widehat{G}^z$ can even be extracted from a finite dimensional transfer matrix. As explained in Section~\ref{sec-NumProc} this suggests a numerical technique to calculate $\Ch_d(P)$ which for $d=2$ was successfully implemented in \cite{ASV} and much more extensively in \cite{AEG}. For $d=4$, for example, the numerical procedure allows to calculate a second Chern number from a four-dimensional system as a three-dimensional winding number. (It is, however, likely that this numerical approach is less efficient than the technique of the spectral localizer \cite{LS}.) From a theoretical perspective, Theorem~\ref{theo-DimReduction} is of interest because it allows to define an effective chiral Hamiltonian $h_{\mbox{\rm\tiny eff}}$ on the doubled boundary Hilbert space
$$
h_{\mbox{\rm\tiny eff}}
\;=\;
\begin{pmatrix}
0 & (\widehat{V}^z)^* \\ \widehat{V}^z & 0 
\end{pmatrix}
\;,
$$
which has the same strong topological invariant as $H$. It hence provides an algorithmic procedure for the dimensional reduction within the class of models considered, hence realizing \cite{QHZ,EG,FHA} in the present framework. A dual result expressing the strong invariants of chiral bulk systems in odd dimensions (namely higher winding numbers) in terms of a selfadjoint gapped operator on the boundary will be presented elsewhere.

\vspace{.2cm}

The second contribution of this paper concerns a scattering theory formulation for the invariants. For this purpose, one considers a set of identical wires which are perfectly conducting at the Fermi energy and couples them to the half-space insulator. As the insulator has no conducting states at this energy, the scattering matrix only consists of a unitary reflection matrix. It is not the object of the present paper to study this set-up within the framework of mathematical scattering theory, but we rather give an adhoc derivation of a formula for the reflection matrix which, when properly defined as in Section~\ref{sec-ScatStates}, is  a covariant operator on the Hilbert space over the boundary $\ZM^{d-1}\times\{1\}$. Once this is carefully spelled out, the following result is merely a corollary of Theorem~\ref{theo-DimReduction}. 

\begin{theo}
\label{theo-ScatInv} 
Let $d$, $H$, $P$, $\widehat{H}$, $\mu$ and $\delta$ be as in {\rm Theorem~\ref{theo-DimReduction}}.  Let $R^{\mu+\imath\delta}$ be the reflection matrix near $\mu$ for the half-space Hamiltonian $\widehat{H}$ coupled to a half-sided conducting wires. Then
$$
\Ch_d(P)
\;=\;
-\,\Ch_{d-1}(R^{\mu+\imath\delta})
\;.
$$ 
\end{theo}

The proof of Theorem~\ref{theo-ScatInv} is given in Section~\ref{sec-ReflMatrix}. For models in continuous physical space and $d=2$ this result was already obtained by Br\"aunlich, Graf and Ortelli \cite{BGO}. The higher dimensional cases give a precise formulation to the suggestive discussion by Fulga, Hassler and Akhmerov \cite{FHA}. Let us note that it is also possible to consider the scattering on a finite size sample of the insulator with wires attached to both sides \cite{BGO}. Then the reflection matrix is not unitary for real energies, but for sufficiently long samples it is still invertible and contains the same topological information. Let us also mention that Levinson's theorem is another element of scattering theory for which $K$-theoretic methods have been implemented sucessfully \cite{KR,BS}. In an extension of this approach \cite{SB4}, covariant scattering matrices on a hypersurface similar to the above play a central role. More precisly, the non-commutative winding of the scattering matrix in the energy variable is shown to be equal to the surface density of states.

\vspace{.2cm}

\noindent {\bf Notations:} We essentially follow the notations and the terminology of the monograph \cite{PS}. In particular, \cite{PS} also contains a careful definition of the operator algebras as well as the $K$-groups and the connecting maps of $K$-theory. Nevertheless, the main results of this work are of analytic nature and  strictly speaking do not concern $K$-theory, but rather homotopy theory. More precisely, a new representative for the image of the exponential map is given. When we write an equality $[U]_1=[V]_1$ in a $K_1$-group it means that there exists a homotopy $U\sim V$ within the invertible operators (possibly enlarged by a matrix degree of freedom). This is essentially sufficient to follow the arguments below.

\section{Dimensional reduction in topological insulators}
\label{sec-DimRedu}

\subsection{Description of the insulator and its strong invariants}
\label{sec-Insulator}

Theorem~\ref{theo-DimReduction} requires $H$ to be a covariant family of nearest neighbor Hamiltonians on $\ell^2(\ZM^d,\CM^L)$ with $d$ even. To give a definition of these notions, let us sketch the algebraic framework for such random operators as developed by Bellissard \cite{Bel} and described in \cite{PS} where full details can be found. Let $(\Omega,T,\ZM^d,\PM)$ be a compact probability space equipped with a $\ZM^d$ action $T$ with respect to which the probability measure $\PM$ is invariant and ergodic. Then a family $A=(A_\omega)_{\omega\in\Omega}$ of operators $A_\omega$  each acting on $\ell^2(\ZM^d,\CM^L)$ is called covariant if it satisfies the covariance relation 
$$
U(a)A_\omega U(a)^*\;=\;A_{T^a\omega}
\;,
\qquad
a\in\ZM^d
\;,
$$ 
for the translations $U(a)$ on $\ell^2(\ZM^d)$. A constant magnetic field can be dealt with by using the magnetic translations instead. An operator $A_\omega$ is called short range if there exists an $R$ such that its matrix elements $\pi_n A_\omega \pi_m^*$ between sites $n\in\ZM^d$ and $m\in\ZM^d$ vanish for $|n-m|>R$. Here $\pi_n$ is the partial isometry from $\ell^2(\ZM^d,\CM^L)$ onto the fiber Hilbert space $\CM^L$ over the site $n$, so that $\pi_n A_\omega \pi_m^*$ is an $L\times L$ matrix. The set of all covariant families of short ranged operators generates a C$^*$-algebra denoted by $\Aa_d$.  This algebra has a  non-commutative differential structure given by the (densely defined) derivations induced by the position operators $ X_j $ on $\ell^2(\ZM^d)$:
$$
\partial_j A_\omega\;=\;\imath[A_\omega, X_j]
\;,
\qquad
j=1,\ldots,d
\;,
$$
and a non-commutative integration fixed by the invariant and ergodic probability measure $\PM$:
$$
\Tt(A)
\;=\;
\int \PM(d\omega)
\;\Tr_L
\left(
\pi_0
A_\omega \pi_0^*
\right)
\;.
$$
Here $ \Tr_L $ denotes the trace over the fiber $\CM^L$. If now $P=(P_\omega)_{\omega\in\Omega}$  is a given (smooth) projection in $\Aa_d$ specifying a class in the group $K_0(\Aa_d)$, its $d$th even Chern number is defined by
\begin{equation}
\label{eq-ChernFinite3}
\Ch_d(P)
\;=\;
\frac{(2\pi\imath)^{\frac{d}{2}}}{\frac{d}{2}!}
\sum_{\sigma\in S_d}(-1)^\sigma 
\;\Tt\left(
P\partial_{\sigma(1)}P\cdots \partial_{\sigma(d)}P
\right)
\;,
\end{equation}
where $S_d$ denotes the symmetric group of permutations over $d$ points and $(-1)^\sigma$ is the signature of the permutation $\sigma$.  This definition applies to the Fermi projection $P=\chi(H\leq\mu)$ if the Fermi level $\mu$ lies in a gap of the ($\PM$-almost sure) spectrum of $H$, and then provides the quantity in Theorems~\ref{theo-DimReduction} and \ref{theo-ScatInv}.

\vspace{.2cm}

It remains to explain the nearest neighbor property of the Hamiltonian. It simply states that the range is $R=1$. This implies that $H_\omega$ can be written in a block Jacobi form
\begin{equation}
\label{eq-threetermrec}
(H_\omega\phi)_n
\;=\;
A_{n+1}\,\phi_{n+1}\,+B_n\,\phi_n
+\,A_n^*\,\phi_{n-1}
\;,
\qquad
n\in\ZM
\;.
\end{equation}
Here $\phi=(\phi_n)_{n\in\ZM}$ with $\phi_n\in \ell^2(\ZM^{d-1},\CM^L)$, $B_n=B_n^*$ are self-adjoint operators on $\ell^2(\ZM^{d-1},\CM^L)$, and $A_n$ are invertible operators on $\ell^2(\ZM^{d-1},\CM^L)$. Throughout it will be supposed that $A_n$, $A_n^{-1}$ and $B_n$ are uniformly (in $\omega$) bounded operators. The dependence on $\omega$ is suppressed unless it is of relevance. It is well-known that the spectrum $\sigma(H)$ is $\PM$-almost surely independent of $\omega$. Furthermore, it is assumed that $H$ describes an insulator in the sense that the Fermi energy $\mu$ lies in a bulk gap $\Delta\subset\RM$ of the spectrum of $H$.

\subsection{The bulk-boundary correspondence}
\label{sec-BBC}

A boundary is introduced by restricting the operators to a half-space which is chosen to be $\ZM^{d-1}\times\NM\subset \ZM^d$, as in \cite{PS}. The restrictions of covariant operator families from $\Aa_d$ generate a C$^*$-algebra $\widehat{\Aa}_d$ of half-space operators. It contains an algebra $\Ee_d$ of operators which are covariant along the boundary and fall off away from the boundary. This algebra is isomorphic to $\Aa_{d-1}\otimes \Kk$ where $\Kk$ are the compact operators on $\ell^2(\NM)$ and this leads to a short exact sequence  of C$^*$-algebras
\begin{equation}
\label{eq-BBCExSeq}
0
\;\to\;
\Ee_d
\;\to\;
\widehat{\Aa}_d
\;\to\;
\Aa_d
\;\to\;
0
\;.
\end{equation}
On $\Ee_d$ the non-commutative derivatives $\partial_1,\ldots,\partial_{d-1}$ are defined by the same formulas, and the trace integration $\Tt$ (on $\Aa_{d-1}$) is extended by the usual trace on $\Kk$ to a (densely defined) trace $\widehat{\Tt}$. Elements of the group $K_1(\Ee_d)$ are represented by an invertible operator $\widehat{U}$ in the unitization $\Ee_d^\sim$ of $\Ee_d$ (or matrices with entries in $\Ee_d^\sim$). For such an operator satisfying, moreover, differentiability and traceclass properties, the $(d-1)$th odd Chern number is defined by (recall that $d$ is even):
\begin{equation}
\label{eq-OddChern}
\Ch_{d-1}(\widehat{U})
\;=\;
\frac{\imath(\imath\pi)^{\frac{d}{2}-1}}{(d-1)!!}
\sum_{\sigma\in S_{d-1}}(-1)^\sigma 
\;\widehat{\Tt}
\left(
\prod_{j=1}^{d-1}\big(
(\widehat{U}^{-1}-\one)\partial_{\sigma(j)}\widehat{U}
\big)
\right)
\;.
\end{equation}
The subtraction of the identity in $\widehat{U}^{-1}-\one$ is done to insure the traceclass property. Of particular interest for the bulk-boundary correspondence is the image $\Exp[P]_0\in K_1(\Ee_d)$  of a class $[P]_0\in K_0(\Aa_d)$ under the exponential map $\Exp$ of $K$-theory described below. 

\begin{theo}[\cite{KRS}, Theorem 5.3.3(i) in \cite{PS}]
\label{theo-BBC} Let $d$ be even. For a differentiable projection $P\in\Aa_d$ and a differentiable invertible $\widehat{U}\in\Ee_d^\sim$ such that $[\widehat{U}]_1=\Exp[P]_0$,
$$
\Ch_d(P)
\;=\;
\Ch_{d-1}(\widehat{U})
\;.
$$ 
\end{theo}

To put this theorem to work, one needs a representative $\widehat{U}$ representing $\Exp[P]_0$ in the case where $P=\chi(H\leq \mu)$ is the Fermi projection. This can be done in terms of a half-space Hamiltonian $\widehat{H}\in\widehat{\Aa}_d$ obtained by restricting $H$ to the half-space \cite{KRS,PS}:
\begin{equation}
\label{eq-ExpDef}
\widehat{U}
\;=\;
e^{2\pi\imath \,f_{\mbox{\rm\tiny Exp}}(\widehat{H})}
\;.
\end{equation}
Here $f_{\mbox{\rm\tiny Exp}}:\RM\to[0,1]$ is a smooth non-decreasing function equal to $0$ below and $1$ above the insulating bulk gap of $ H $ containing $\mu$. In particular, $\Ch_{d-1}(\widehat{U})$ can be shown to be well-defined. The aim of the following is to find different representatives for the class $[\widehat{U}]_1$ which only have non-vanishing matrix entries in a finite strip away from the boundary and eventually only on the boundary itself.

\subsection{Strict boundary formulation of boundary invariants}
\label{sec-strict}

The operator $\widehat{U}-\one$ given in \eqref{eq-ExpDef} is obtained by functional calculus from $\widehat{H}$ with a function supported on a bulk gap (of $H$). This is known to imply a decay away from the boundary. More precisely, for $n\in\NM$ let $\Pi_n:\ell^2(\ZM^{d-1}\times\NM,\CM^L) \rightarrow  \ell^2(\ZM^{d-1}\times\{n\},\CM^L)$ be the partial isometry onto $\ell^2(\ZM^{d-1}\times\{n\},\CM^L)\cong\ell^2(\ZM^{d-1},\CM^L)$, namely onto the Hilbert space over the sites $\ZM^{d-1}\times\{n\}$ of the half-space $\ZM^{d-1}\times\NM$. Note the difference with $ \pi_n $ used above acting on $\ell^2(\ZM^{d},\CM^L)$. Then  ({\sl e.g.} \cite{PS})
$$
\|\Pi_n\widehat{U}\,\Pi_m^*\|
\;\leq\;
C\,e^{-\beta (n+m)}
\;,
\qquad n\not = m\;,
$$
with $\beta>0$ and $C$ which can be chosen uniformly in $\omega$. However, none of the restrictions $\Pi_n\widehat{U}\Pi_m^*$ vanishes identically. Setting $\Pi_{[1,N]}=(\Pi_1 , \ldots, \Pi_N) $ with range $\ell^2(\ZM^{d-1}\times\{1,\ldots,N\},\CM^L)$, one therefore obtains that
$$
\widehat{U}
\;=\;
\begin{pmatrix}
\Pi_{[1,N]}
\widehat{U}\,
\Pi_{[1,N]}^* & 0 \\
0 &
\one
\end{pmatrix}
\;+\;
K_N
\;,
$$
where $K_N\in\Ee_d$ is an operator satisfying $\lim_{N\to\infty}\|K_N\|=0$. In particular, for $N$ large enough $\Pi_{[1,N]}\widehat{U}\,\Pi_{[1,N]}^*$ is invertible. Hence one can homotopy $K_N$ down to $0$ without violating the invertibility so that
$$
[\widehat{U}]_1
\;=\;
\left[
\Pi_{[1,N]}
\widehat{U}\,
\Pi_{[1,N]}^*
\right]_1
\;\in\;K_1(\Ee_d)
\;.
$$
Now the representative on the r.h.s. is by construction supported only on a strip of width $N$ off the boundary. Let us call this a strict boundary representative of the class $[\widehat{U}]_1$ that will lead to a strict boundary formulation of the invariant. Its Chern number can now be calculated by \eqref{eq-OddChern}, or alternatively without the subtraction of the identity, namely for a (smooth) invertible $\widehat{V}\in\Ee_d$ supported on a finite strip (such as the operator $\Pi_{[1,N]}\widehat{U}\,\Pi_{[1,N]}^*$),
\begin{equation}
\label{eq-OddChern2}
\Ch_{d-1}(\widehat{V})
\;=\;
\frac{\imath(\imath\pi)^{\frac{d}{2}-1}}{(d-1)!!}
\sum_{\sigma\in S_{d-1}}(-1)^\sigma 
\;\widehat{\Tt}
\left(
\prod_{j=1}^{d-1}\big(
\widehat{V}^{-1}\partial_{\sigma(j)}\widehat{V}
\big)
\right)
\;.
\end{equation}
The next step is to show that one can use a suitable function of the restriction of the resolvent of $\widehat{H}$ instead of the restriction of $\widehat{U}$. For $z=\mu+\imath\delta\in\CM$ let us set
$$
\widehat{G}^z(n,m)
\;=\;
\Pi_n(\widehat{H}\,-\,z\, \one)^{-1}\Pi_m^*
\;,
$$
and
$$
\widehat{G}^z_N
\;=\;
\big(\widehat{G}^z(n,m)\big)_{n,m=1,\ldots,N}
\;=\;
\Pi_{[1,N]}
(\widehat{H}\,-\,z\,\one )^{-1}
\Pi_{[1,N]}^*
\;.
$$
The next result shows that a Cayley transform of $\widehat{G}^z_N$ represents $[\widehat{U}]_1$ and hence provides a new way to express the image of the $K$-theoretic exponential map in terms of resolvents. The reader familiar with the connecting maps of $K$-theory of C$^*$-algebras readily sees that this fact and the result leading to it is not restricted to the exact sequence \eqref{eq-BBCExSeq}, but rather extends to a much larger class of exact sequences. On the other hand, Theorem~\ref{theo-DimReduction} goes further and requires supplementary hypotheses and analytical arguments.

\begin{theo}
\label{theo-FiniteGreen} 
For any even $d$ and $\delta>0$ sufficiently small,
$$
[\widehat{U}]_1
\;=\;
-\,\left[(2 \delta\, \widehat{G}^{\mu+\imath \delta}_N\,-\,\imath\,\one_N)(2 \delta\, \widehat{G}^{\mu+\imath \delta}_N\,+\,\imath\,\one_N )^{-1}
\right]_1
\;,
$$
where $\one_N=\Pi_{[1,N]}\Pi_{[1,N]}^*$.
\end{theo}

\noindent {\bf Proof:}
Let us introduce two $\SM^1$-valued functions $ F $ and $ F_{\delta} $ on $\RM$ by
\begin{equation}
\label{eq-Functions}
F(E)\;=\;
e^{-2\pi\imath \,f_{\mbox{\rm\tiny Exp}}(E)}
\;,
\qquad
F_{\delta}(E)
\;=\;
\frac{E-\mu+\imath {\delta}}{E-\mu-\imath {\delta}}
\;=\;
1\,+\,\frac{2\imath {\delta}}{E-\mu-\imath {\delta}}
\;,
\end{equation}
where $ f_{\mbox{\rm\tiny Exp}} $ has been defined before \eqref{eq-ExpDef}. Both functions are  close to $1$ for large $E$ and make one negatively oriented loop, namely have a winding number equal to $-1$. Moreover, $\lim_{{\delta}\to 0}F_{\delta}(E)= 1$ for $E\not=\mu$. The parameter $ \delta $ is chosen small enough so that $F_{\delta} $ is close to $ 1 $ outside of a neighborhood of $ \mu $.  There is still the freedom of choosing the non-decreasing function $f_{\mbox{\rm\tiny Exp}}$ with derivative supported in the bulk gap $\Delta\subset\RM$ containing $\mu$. It will be chosen such that $F$ coincides with $F_{\delta}$ on an interval $[\mu-{\delta}^{\frac{1}{2}},\mu+{\delta}^{\frac{1}{2}}]$. By smoothly completing the function $F$ and thus also $f_{\mbox{\rm\tiny Exp}}$, one can assure
$$
\|F-F_{\delta}\|_\infty\;\leq\;C{\delta}^{\frac{1}{2}}
\;,
$$
for a suitable constant $C$. As $\widehat{U}^*=F(\widehat{H})$ by \eqref{eq-ExpDef}, one concludes
$$
\left\|\widehat{U}^*\,-\,F_{\delta}(\widehat{H})\right\|
\;\leq\;
C{\delta}^{\frac{1}{2}}\;.
$$
For ${\delta}$ sufficiently small, the invertibility of the finite-volume restrictions is not violated and one concludes that
$$
[\widehat{U}^*]_1
\;=\;
\big[
\Pi_{[1,N]}F_{\delta}(\widehat{H})
\Pi_{[1,N]}^*
\big]_1
\;=\;
\big[
\one_N\,+\,2\imath {\delta}\, \widehat{G}^{\mu+\imath {\delta}}_N
\big]_1
\;=\;
\big[
2{\delta}\, \widehat{G}^{\mu+\imath {\delta}}_N\,-\,\imath\,\one_N
\big]_1
\;,
$$
where the second equality of \eqref{eq-Functions} was used. Note that this implies, in particular, that $2 {\delta}\, \widehat{G}^{\mu+\imath {\delta}}_N-\imath\,\one_N$ is invertible. On the other hand, as $2 {\delta}\, \widehat{G}^{\mu+\imath {\delta}}_N+\imath\,\one_N$ has a strictly positive imaginary part (for ${\delta}>0$), it can be homotopically deformed into the identity and is therefore trivial in $K_1(\Ee_d)$. This concludes the proof.
\hfill $\Box$

\subsection{Limit behavior of the Cayley transform of the resolvent}
\label{sec-LimitCayley}

Due to Theorem~\ref{theo-FiniteGreen}, one is led to study the operator
$$
\widehat{V}^z_{N,\epsilon}
\;=\;
(2 \epsilon\, \widehat{G}^{z}_N\,-\,\imath\,\one_N)(2 \epsilon\, \widehat{G}^{z}_N\,+\,\imath\,\one_N )^{-1}
\;.
$$
For $z=\mu+\imath \delta$, $\epsilon=\delta$ and $\delta>0$ sufficiently small it represents $[\widehat{U}^*]_1\in K_1(\Ee_d)$.  As the imaginary part of $\widehat{G}^z_{N}$, as usual defined by $\Im m(G)=(2\imath)^{-1}(G-G^*)$, has the same sign of the imaginary part of $z$, it follows that the operator $\widehat{G}^z_N$ lies in the upper or lower half-plane and is thus invertible for $\Im m(z)\not=0$ (see Appendix~\ref{app-Cayley}). Therefore one can also write
\begin{equation}
\label{eq-VGinv}
\widehat{V}^z_{N,\epsilon}
\;=\;
\big(2 \epsilon\imath\,\one_N\,+\, (\widehat{G}^{z}_N)^{-1}\big)\big(2 \epsilon\imath\,\one_N\,-\, (\widehat{G}^{z}_N)^{-1}\big)^{-1}
\;.
\end{equation}
Another implication of $\Im m(\widehat{G}^z_N)>0$, following from Proposition~\ref{prop-Cayley} in Appendix~\ref{app-Cayley}, is that
\begin{equation}
\label{eq-CayleyBound}
\|\widehat{V}^z_{N,\epsilon}\|\,\leq\, 1\;,
\qquad
\Im m(z)>0
\;.
\end{equation}
Hence $z\in\HM\mapsto \widehat{V}^z_{N,\epsilon}$ is an analytic map from the upper half-plane $\HM$ to the unit disc $\DM(\Hh)\subset\Bb(\Hh)$ of operators as defined in Appendix~\ref{app-Cayley}. It is a classical result that such functions have non-tangential limit points 
\begin{equation}
\label{eq-LimitExists}
\lim_{\delta\downarrow 0}\;\widehat{V}^{\mu+\imath\delta}_{N,\epsilon}
\end{equation}
for Lebesgue almost all $\mu$, at least in the weak sense, {\it e.g.} Theorem~11.20 in \cite{Rud}. For periodic operators along the boundary,  we will show convergence in norm and to a unitary operator.  Note that if there were no spectrum of $\widehat{H}$ at $\mu$, then the Green matrix would have no imaginary part and the limit would thus indeed be unitary. This corresponds to a trivial insulator though, and will not be considered here. For a topological insulator, there is spectrum of $\widehat{H}$ at $\mu$ corresponding to boundary states and this does lead to a non-vanishing imaginary part of the Green matrix. However, as such states are of codimension $1$ (in particular, they vanish in the density of states due to the bulk gap), one may expect the resulting singularities to be removable. This is what is proven in the following.  

\vspace{.2cm}

If $\widehat{H}$ is periodic in the $(d-1)$ directions tangential to the boundary, it can be partially diagonalized by the Bloch-Floquet transformation:
\begin{equation}
\label{eq-HHastDiag}
\widehat{H}
\;\cong\;
\int^\oplus_{\TM^{d-1}} de^{\imath k}\;\widehat{H}(e^{\imath k})
\;,
\end{equation}
where $\widehat{H}(e^{\imath k})$ is a half-sided block Jacobi operator on $\ell^2(\NM,\CM^L)$. As $\widehat{H}$ is supposed to be of finite range, it follows that $\widehat{H}(e^{\imath k})$ can be seen as an analytic function in $\zeta=e^{\imath k}$ which extends to an operator valued analytic function $\widehat{H}(\zeta)$ on $\TM^{d-1}_\delta=\{\zeta\in\CM^{d-1}\,:\,d(\zeta,\TM^{d-1})<\delta\}$ where $d$ is the euclidean distance. It satisfies
$$
\widehat{H}(\zeta)^*\;=\;\widehat{H}\big((\overline{\zeta})^{-1}\big)
\;.
$$
Then $\widehat{G}^z_N(\zeta)=\Pi_{[1,N]}(\widehat{H}(\zeta)-z\one)^{-1}\Pi_{[1,N]}^*$ is a $NL\times NL$ matrix which is analytic in $z\in\CM\setminus\RM$ and $\zeta\in\TM^{d-1}_\delta$  satisfying
$$
\widehat{G}^z_N(\zeta)^*
\;=\;
\widehat{G}^{\overline{z}}_N\big((\overline{\zeta})^{-1}\big)
\;.
$$
Moreover, this function extends to the real axis for many values of $\zeta$. Recall that as $\mu$ lies outside of the essential spectrum of $H(e^{\imath k})$, it does so also for $\widehat{H}(e^{\imath k})$. Hence $\widehat{H}(e^{\imath k})$ has only discrete spectrum of finite multiplicity in a complex ball $B_\delta(\mu)\subset\CM$ around $\mu$ and, of course, these eigenvalues lie on the real axis for $\zeta=e^{\imath k}\in\TM^{d-1}$. For such a $\zeta=e^{\imath k}\in\TM^{d-1}$ and $\Im m(z)\not=0$, let us now consider
\begin{equation}
\label{eq-VNdef}
\widehat{V}^z_{N,\epsilon}(\zeta)
\;=\;
\big(2\epsilon\,\widehat{G}^z_N(\zeta)-\imath\,\one_N\big)\big(2\epsilon\,
\widehat{G}^z_N(\zeta)+\imath\,\one_N\big)^{-1}
\;,
\end{equation}
for which
\begin{equation}
\label{eq-VNAdj}
\widehat{V}^z_{N,\epsilon}(\zeta)^*
\;=\;
\big(\widehat{V}^{\overline{z}}_{N,\epsilon}(\zeta)\big)^{-1}
\;.
\end{equation}
Except for a discrete set of singularities given the eigenvalues of $\widehat{H}(\zeta)$, this function extends to all $z$ in $(\mu-\delta,\mu+\delta)$ as a function with values in the unitary matrices.  Due to \eqref{eq-CayleyBound}, $\widehat{V}^z_{N,\epsilon}(\zeta)$ is bounded by $1$ for $\Im m(z)>0$. Another important point is that the function $z\mapsto \widehat{V}^z_{N,\epsilon}(\zeta)$ is meromorphic (notably without essential singularities) on a strip with $\Re e(z) \in (\mu-\delta,\mu+\delta)$ because $z\mapsto \widehat{G}^z_N(\zeta)$ has only poles and $\widehat{V}^z_{N,\epsilon}(\zeta)$ is a rational function of $\widehat{G}^z_N(\zeta)$. This implies that also $z\mapsto \widehat{V}^z_{N,\epsilon}(\zeta)^{-1}$ meromorphic on this strip since each of its entries is given by a product of $\det(\widehat{V}^z_{N,\epsilon}(\zeta))^{-1}$ and the determinant of a minor of $\widehat{V}^z_{N,\epsilon}(\zeta)$. Moreover, as long as  $z\in(\mu-\delta,\mu+\delta)$ is not an eigenvalue of $\widehat{H}(\zeta)$, one has $\det(\widehat{V}^z_{N,\epsilon}(\zeta))\in\SM^1$ so that $\det(\widehat{V}^z_{N,\epsilon}(\zeta))$ has no zeros on $z\in(\mu-\delta,\mu+\delta)$. It follows that there exists a $\delta_0'>0$ such that $\widehat{V}^z_{N,\epsilon}(\zeta)^{-1}$ is also a uniformly bounded function on a rectangle $K_{\delta_0,\delta'_0}=\{z\in\CM\,:\,|\Re e(z)-\mu|\leq \delta_0,\;0<\Im m(z)\leq \delta'_0\}$. Hence for $z\in K_{\delta_0,\delta'_0}$ and still $\zeta=e^{\imath k}\in\TM^{d-1}$,  \eqref{eq-VNAdj} implies that also $\|\widehat{V}^{\overline{z}}_{N,\epsilon}(\zeta)\|\leq C$ for some $C>0$. Together, $z\in  K_{\delta_0,\delta'_0}\cup \overline{ K_{\delta_0,\delta'_0}}\mapsto \widehat{V}^z_{N,\epsilon}(\zeta)$ is bounded and by the Riemann removable singularity theorem (applied to each matrix entry of $\widehat{V}^z_{N,\epsilon}(\zeta)$), one can remove the finite number of singularities. The continuously extended function is then unitary for all $z\in (\mu-\delta,\mu+\delta)$ and $\zeta=e^{\imath k}\in\TM^{d-1}$.

\vspace{.2cm}

It is furthermore necessary to show that $\widehat{V}^z_{N,\epsilon}(\zeta)$ is differentiable in $\zeta$. This can be shown by applying the Riemann removable singularity theorem in several complex variables ({\it e.g.} Theorem~4.2.1 in \cite{Sch}). Indeed, by analytic perturbation theory the discrete eigenvalues of $\widehat{H}(\zeta)$ are analytic curves in $\zeta$, provided that the branches are labelled in a suitable manner at level crossings. Therefore 
$$
A
\;=\;
\{(z,\zeta)\in B_\delta(\mu)\times\TM^{d-1}_\delta\;:\;z\;\mbox{\rm eigenvalue of }\widehat{H}(\zeta)\}
$$
is an analytic set of complex codimension equal to $1$ in the (standard) sense of \cite{Sch}, unless there is no eigenvalue at all. Having no eigenvalues would mean that the insulator is trivial, a case which is not further investigated here.  Then $(z,\zeta)\in (B_\delta(\mu)\times\TM^{d-1}_\delta)\setminus A\mapsto \widehat{G}^z_N(\zeta)$ is analytic and therefore, for any $\epsilon>0$, so is $(z,\zeta)\in (B_\delta(\mu)\times\TM^{d-1}_\delta)\setminus A \mapsto\widehat{V}^z_{N,\epsilon}(\zeta)$. The task is then to show that this latter function is bounded in a neighborhood of $A$. Here we follow a less ambitious, but sufficient route and only check the analyticity in $\zeta$ for fixed real $z$.

\begin{lemma}
\label{lem-analyticity} 
For $z\in (\mu-\delta,\mu+\delta)$ and $\epsilon>0$, the map $\zeta\in\TM^{d-1}\mapsto \widehat{V}^z_{N,\epsilon}(\zeta)$ is real analytic in each of its components. 
\end{lemma}

\noindent {\bf Proof:} (The argument below is similar to that in the appendix of \cite{BS}.) Because each component is considered separately, it is sufficient to restrict to the case $d=2$. Fix a point $\zeta_0\in\TM^1$ and choose $\delta'$ such that the number of eigenvalue branches of $\widehat{H}(\zeta)$ in $B_\delta(z)$ is constant in a neighborhood $B_{\delta'}(\zeta_0)\subset\CM$. Let $P(\zeta)$ denote the finite-dimensional spectral projection of $\widehat{H}(\zeta)$ onto these eigenvalues and set $Q(\zeta)=\one-P(\zeta)$. By analytic perturbation theory both of these projections are analytic. Then the Green matrix $\widehat{G}^z_N(\zeta)$ can be decomposed in a singular and a regular part. More precisely, let us set
$$
s(\zeta)\;=\;2\epsilon\,\Pi_{[1,N]}P(\zeta)(\widehat{H}(\zeta)-z)^{-1}\Pi_{[1,N]}^*
\;,
\qquad
r(\zeta)\;=\;2\epsilon\,\Pi_{[1,N]}Q(\zeta)(\widehat{H}(\zeta)-z)^{-1}\Pi_{[1,N]}^*
\;.
$$
With these notations, $2\epsilon \,\widehat{G}^z_N(\zeta)=s(\zeta)+r(\zeta)$. Then $\zeta\in B_{\delta'}(\zeta_0)\mapsto r(\zeta)$ is analytic and $\zeta\in B_{\delta'}(\zeta_0)\mapsto s(\zeta)$ is analytic away from singularities that have to be dealt with. Away from these singularities, one can diagonalize $s(\zeta)$ by an analytic basis change $u(\zeta)$:
$$
u(\zeta)\,s(\zeta)\,u(\zeta)^{-1}\;=\;
\begin{pmatrix}
\eta(\zeta) & 0 \\ 0 & 0
\end{pmatrix}
\;,
\qquad
u(\zeta)\,r(\zeta)\,u(\zeta)^{-1}\;=\;
\begin{pmatrix}
a(\zeta) & b(\zeta) \\ c(\zeta) & d(\zeta)
\end{pmatrix}
\;,
$$
where $\eta(\zeta)$ is bounded away from $0$, but has singularities. The block form of $u(\zeta)s(\zeta)u(\zeta)^{-1}$ reflects that $P(\zeta)$ and thus also $s(\zeta)$ are of lower rank. Now replacing into \eqref{eq-VNdef} leads to
$$
\widehat{V}^z_{N,\epsilon}(\zeta)
\;=\;
\big(s(\zeta)+r(\zeta)+\imath\,\one_N\big)^{-1}
\big(s(\zeta)+r(\zeta)-\imath\,\one_N\big)
\;,
$$
so that, suppressing the argument $\zeta$ on $\eta,a,b,d$ for sake of notational simplicity,
$$
u(\zeta)\,\widehat{V}^z_{N,\epsilon}(\zeta)\,u(\zeta)^{-1}
\;=\;
\begin{pmatrix}
\eta + a+\imath & b \\ c & d+\imath
\end{pmatrix}^{-1}
\left[
\begin{pmatrix}
\eta & 0 \\ 0 & 0
\end{pmatrix}
+
\begin{pmatrix}
a-\imath & b \\ c & d-\imath
\end{pmatrix}
\right]
\;.
$$
The first matrix inverse is bounded, and so is the second summand in the bracket.  Indeed, $(\eta+a+\imath)^{-1}$ and $(d+\imath)^{-1}$ exist and have a negative imaginary part for $z\in\RM$ and $\zeta\in\TM^1$ because $\eta$, $a$ and $d$ are then selfadjoint. This guarantees the invertibility which persists for $\zeta\in B_{\delta'}(\zeta_0)$ as long as $\delta'$ is sufficiently small. However, the first summand in the backet has a singularity. To show that it is compensated, let us calculate the matrix inverse using the Schur complement $\sigma=\sigma(\zeta)$: 
$$
\sigma
\;=\; 
d+\imath-c(\eta+a+\imath)^{-1}b
\;.
$$
Again for $\zeta\in\TM^1$, one has $c=b^*$ so that $-c(\eta+a+\imath)^{-1}b$ has a positive imaginary part. Thus $\sigma$ has a positive imaginary part and is thus invertible with uniformly bounded inverse. Due to continuity this also hols on a small neighborhood $B_{\delta'}(\zeta_0)$. By the Schur complement one has
$$
\begin{pmatrix}
\eta + a+\imath & b \\ b^* & d+\imath
\end{pmatrix}^{-1}
\begin{pmatrix}
\eta & 0 \\ 0 & 0
\end{pmatrix}
\;=\;
\begin{pmatrix}
\big(\one+(\eta + a+\imath)^{-1}b\,\sigma^{-1} b^*\big)(\eta + a+\imath)^{-1}\eta & 0 \\ -\sigma^{-1} b^*(\eta + a+\imath)^{-1}\eta  & 0
\end{pmatrix}
\;.
$$
Now not only is $(\eta + a+\imath)^{-1}$ bounded, but also $(\eta + a+\imath)^{-1}\eta=(\one + \eta^{-1}(a+\imath))^{-1}$. Hence the whole matrix is bounded. In conclusion, $u(\zeta)\widehat{V}^z_{N,\epsilon}(\zeta)u(\zeta)^{-1}$ and thus also $\widehat{V}^z_{N,\epsilon}(\zeta)$ are bounded in a neighborhood of the singularities of $\eta$. Consequently these singularities can be removed and $\widehat{V}^z_{N,\epsilon}(\zeta)$ is analytic in $\zeta$.
\hfill $\Box$

\vspace{.2cm}

Due to the smoothness of $\widehat{V}^z_{N,\epsilon}(\zeta)$ in $\zeta$, one can apply the inverse Fourier transform and obtains a differentiable element $\widehat{V}^z_{N,\epsilon}$ in the edge algebra $\Ee_d$. The dependence of $\widehat{V}^z_{N,\epsilon}$ on $\epsilon$ can be analyzed by similar techniques and is also analytic. One concludes that it is possible to deform $\epsilon$ within the positive reals, {\it e.g.} to $\epsilon=\frac{1}{2}$, and then set 
$$
\widehat{V}^z_N
\;=\;
\widehat{V}^z_{N,\frac{1}{2}}
\;=\;
\big(\widehat{G}^z_N-\imath\,\one_N\big)\big(
\widehat{G}^z_N+\imath\,\one_N\big)^{-1}
\;.
$$
One then has 
\begin{equation}
\label{eq-ExpIntermed}
[\widehat{U}]_1
\;=\;
-\,[\widehat{V}^z_N]_1
\;,
\end{equation}
as long as $\Im m(z)$ is sufficiently small so that the $\widehat{V}^z_N$ remains invertible. In particular, $\Im m(z)=0$ is allowed for which $\widehat{V}^z_N$ is unitary. The invertibility persists under weak (possibly random) perturbations.

\begin{proposi}
\label{prop-CayleyInvertible} 
Let $H=H_0+\lambda H_1\in\Aa_d$ with $H_0$ periodic in the $d-1$ directions along the boundary and let $\mu\not\in\sigma(H)$. For $\delta>0$ and $\lambda$ sufficiently small, $\widehat{V}^{\mu+\imath\delta}_N$ is invertible and
$$
[\widehat{U}]_1
\;=\;
-\,[\widehat{V}^{\mu+\imath\delta}_N]_1
\;.
$$
\end{proposi}

\noindent {\bf Proof:} Denote $\widehat{G}^{z,\lambda}_N=\Pi_{[1,N]}
(\widehat{H}_0+\lambda \widehat{H}_1-z \one)^{-1}\Pi_{[1,N]}^*$. The invertibility of $\widehat{G}^{z,\lambda}_N+\imath\,\one_{NL}$ is guaranteed due to a positive imaginary part. It hence has to be shown that $\widehat{G}^{z,\lambda}_N-\imath\,\one_{NL}$ remains invertible. By the resolvent identity,
$$
\widehat{G}^{z,\lambda}_N-\imath\,\one_{NL}
\;=\;
\widehat{G}^{z,0}_N-\imath\,\one_{NL}
\;-\;
\lambda\,
\Pi_{[1,N]}(\widehat{H}_0-z \one)^{-1}\widehat{H}_1
(\widehat{H}_0+\lambda \widehat{H}_1-z\one )^{-1}\Pi_{[1,N]}^*
\;.
$$
Now one first chooses $\delta>0$ sufficiently small so that $\widehat{G}^{z,0}_N-\imath\,\one_N$ is invertible. Then the second summand can be bounded by a constant times $\lambda\delta^{-2}$. Choosing $\lambda$ sufficiently small the second summand thus does not violate the invertibility of $\widehat{G}^{z,\lambda}_N-\imath\,\one_N$.
\hfill $\Box$

\vspace{.2cm}

Let us conclude this section with a few comments. Clearly, it is unsatisfactory to prove the existence of the limit \eqref{eq-LimitExists} only in  the perturbative regime considered Proposition~\ref{prop-CayleyInvertible}. On the other hand, it is precisely the same regime for which Mourre estimates allow to show that the edge spectrum is absolutely continuous \cite{BP,FGW}. Furthermore, going through the above arguments in a quantitative way shows that the actual values of $\delta$ and $\lambda$ for which the estimates hold are not so small after all. Let us also note that it is an interesting open question to analyze the fate of the limit in a mobility gap regime for the bulk Hamiltonian. 

\vspace{.2cm}

Finally let us comment on the parameter $k\in\TM^{d-1}$. In the presentation above, it stems from a partial Bloch-Floquet transform of a $d$ dimensional insulator that is periodic in the $d-1$ dimensions along the boundary. Alternatively, it can simply be the parameters of an external driving of a given setting. For the latter, the most prominent case is $d=2$ corresponding to a time-periodic driving, see \cite{Tho,BGO,PS}. But it is also conceivable that the driving parameters lie in another $d-1$ dimensional parameter space, like the sphere $\SM^{d-1}$. This case is also covered by the analysis in the paper, modulo modifications of the bulk-boundary exact sequence. 

\subsection{Image of the exponential map on the boundary}
\label{sec-ExpImage}

It now remains to show that $\widehat{V}^z_N$ in \eqref{eq-ExpIntermed} can be replaced by $\widehat{V}^z=\widehat{V}^z_1$. Combined with the bulk-boundary correspondence as given in Theorem~\ref{theo-BBC} this concludes the proof of Theorem~\ref{theo-DimReduction}.

\begin{theo}
\label{theo-ExpImage} Let $d$ be even and $H=H_0+\lambda H_1\in\Aa_d$ with $H_0$ periodic in the $d-1$ directions along the boundary, $\delta>0$ and $\lambda$ sufficiently small so that {\rm Proposition~\ref{prop-CayleyInvertible}} applies for $N=1$. Furthermore, let $P=\chi(H\leq\mu)\in\Aa_d$ be the Fermi projection below $\mu\not\in\sigma(H)$. Then
$$
\Exp[P]_0
\;=\;
-\,[(\widehat{G}^{\mu+\imath\delta}\,-\,\imath\,\one_L)(\widehat{G}^{\mu+\imath\delta}\,+\,\imath\,\one_L)^{-1}]_1
\;.
$$
\end{theo}

The proof is based on further homotopy arguments combined with the Schur complement formula and transfer matrix methods. We recall these standard tools in the proof.

\vspace{.2cm}

\noindent {\bf Proof.} First of all, let us note that it is sufficient to consider the case $\lambda=0$, namely that of a partially periodic Hamiltonian $H=H_0$. Indeed, the argument in the proof of Proposition~\ref{prop-CayleyInvertible} allows to deform $\widehat{V}^{\mu+\imath\delta}_N$ with $\lambda$ without violating the invertibility, for all $N\geq 1$. Hence one can assume from now on that $H\cong \int^\oplus_{\TM^{d-1}}de^{\imath k} \,H(e^{\imath k})$. This simplifies the arguments below because one can assume the energy to be on the real axis where $E\mapsto \widehat{V}^E_N$ is unitary, after the removal of the singularities explained in Section~\ref{sec-strict}. In the following, all objects hence depend on $e^{\imath k}\in\TM^{d-1}$ after a basis change, but this will be suppressed in the notations.

\vspace{.1cm}

The proof starts from \eqref{eq-ExpIntermed} with $\widehat{V}^z_N$ given in the form \eqref{eq-VGinv} with $\epsilon=\frac{1}{2}$.  It is now useful to express $\widehat{G}^z_N$ using the Schur complement formula corresponding to the splitting 
\begin{equation}
\label{eq-grading}
\ell^2(\ZM^{d-1}\times \NM,\CM^L)
\;=\;
\ell^2(\ZM^{d-1}\times \{1,\ldots,N\},\CM^L)\oplus \ell^2(\ZM^{d-1}\times \{N+1,N+2,\ldots\},\CM^L)
\;,
\end{equation}
of the Hilbert space, for which
$$
\widehat{H}
\;=\;
\begin{pmatrix}
H_N & \Pi_N^* A_{N+1} \Pi_{N+1} \\ \Pi_{N+1}^* A_{N+1}^*\Pi_N & \widehat{H}_{N+1}
\end{pmatrix}
\;.
$$
Here $H_N$ is a finite size block Jacobi matrix and $\widehat{H}_{N+1}$ a semi-infinite one for which we set
$$
\widetilde{G}^z_{N+1}
\;=\;
\Pi_{N+1}(\widehat{H}_{N+1}-z\,\one)^{-1}\Pi_{N+1}^*
\;.
$$
Then by the Schur complement formula
$$
\widehat{G}^z_N
\;=\;
\left(H_N-z\,\one-\Pi_N^* A_{N+1} \widetilde{G}^z_{N+1}A_{N+1}^*\Pi_N \right)^{-1}
\;.
$$
Replacing in \eqref{eq-VGinv} shows
$$
\widehat{V}^z_N
=
\left(H_N-z\,\one-\pi_N^* A_{N+1} \widetilde{G}^z_{N+1}A_{N+1}^*\pi_N+\imath\,\one \right)
\left(H_N-z\,\one-\pi_N^* A_{N+1} \widetilde{G}^z_{N+1}A_{N+1}^*\pi_N-\imath \,\one\right)^{-1}
\!\!.
$$
Now for $\delta=\Im m(z)=0$, this is the Cayley transform of a selfadjoint operator (with removable singularities stemming from $\widetilde{G}^z_{N+1}$) and therefore it is unitary. As $H_N-z$ is then selfadjoint it can be homotopically turned down to $0$ without violating the unitarity, and thus in particular the invertibility. Hence $\widehat{V}^z_N$ can be homotopically deformed as follows 
\begin{align*}
{\widehat{V}}^z_N
&
\;\sim\;
\left(-\pi_N^* A_{N+1} \widetilde{G}^z_{N+1}A_{N+1}^*\pi_N+\imath \,\one\right)
\left(-\pi_N^* A_{N+1} \widetilde{G}^z_{N+1}A_{N+1}^*\pi_N-\imath \,\one\right)^{-1}
\\
&
\;=\;
\begin{pmatrix}
\one_{L(N-1)} & 0 \\ 0 & 
A_{N+1} (-\widetilde{G}^z_{N+1}+\imath (A_{N+1}^*A_{N+1} )^{-1} )(-\widetilde{G}^z_{N+1}-\imath(A_{N+1}^*A_{N+1} )^{-1})^{-1}A_{N+1}^{-1}
\end{pmatrix}
\\
&
\;\sim\;
\begin{pmatrix}
-\one_{L(N-1)} & 0 \\ 0 & 
A_{N+1} (-\widetilde{G}^z_{N+1}+\imath\,\one)(-\widetilde{G}^z_{N+1}-\imath\,\one)^{-1}A_{N+1}^{-1}
\end{pmatrix}
\;,
\end{align*}
where in the last step $A_{N+1}^{-1}(A_{N+1}^{-1})^*\sim\one$ in the positive operators was used for the homotopies $\widetilde{G}^z_{N+1}\pm \imath (A_{N+1}^*A_{N+1})^{-1}\sim \widetilde{G}^z_{N+1}\pm\imath\,\one$ within the invertibles. Hence
\begin{align*}
[\widehat{V}^z_N]_1
&
\;=\;
\big[A_{N+1} (\widetilde{G}^z_{N+1}-\imath\,\one)(\widetilde{G}^z_{N+1}+\imath\,\one)^{-1}A_{N+1}^{-1}\big]_1
\\
& 
\;=\;
\big[A_{N+1}\big]_1
\,+\,
\big[(\widetilde{G}^z_{N+1}-\imath\,\one)(\widetilde{G}^z_{N+1}+\imath\,\one)^{-1}\big]_1
\,-\,
\big[A_{N+1}\big]_1
\\
& 
\;=\;
\big[\widetilde{V}^z_{N+1}\big]_1
\;,
\end{align*}
where
$$
\widetilde{V}^z_{N+1}
\;=\;
(\widetilde{G}^z_{N+1}-\imath\,\one)(\widetilde{G}^z_{N+1}+\imath\,\one)^{-1}
\;.
$$
Now remains to show that $\widetilde{V}^z_{N+1} $ and $\widehat{V}^z_1$ are homotopic inside the invertible operators. This is accomplished by studying the generalized eigenfunctions for block Jacobi operators ({\it e.g.} \cite{SB2}), applied either for each $k$ separately or extended to operator valued Jacobi operators. The basic fact is that the (formal and not necessarily square integrable in $n$) solutions  $\phi=(\phi_n)_{n\in\ZM}$  of the Schr\"odinger equation $H\phi=z\phi$ at energy $z\in\CM$ can be calculated using the transfer matrices
\begin{equation}
\label{eq-transdef} 
\Tt^z_n
\;=\;
\begin{pmatrix}
(z\,{\bf 1}-B_n)A_n^{-1}
& -\,A_n^*
\\
A_n^{-1} & 0
\end{pmatrix}\;,
\end{equation}
namely for $n\geq 1$ one has
\begin{equation}
\label{eq-transrel}
\Phi^z_n
\;=\;
\Tt^z_n\,
\Phi^z_{n-1}
\;,
\qquad
\mbox{\rm where }\;
\Phi^z_n
\;=\;
\begin{pmatrix}
A_{n+1}\phi_{n+1} \\ \phi_{n}
\end{pmatrix}
\;.
\end{equation}
Here the initial condition $\Phi^z_0$ selects one solution, and $\Tt^z_1$ is calculated with $A_1=\one$ (note that $A_1$ is not part of $\widehat{H}$).  For $z=E\in\RM$, the transfer matrices satisfy
\begin{equation}
\label{eq-GUnitarity}
(\Tt^E_n)^*\Gg\Tt^E_n\;=\;\Gg\;,
\qquad
\Gg \;=\;
\imath \begin{pmatrix}
0 & -\one
\\
\one & 0
\end{pmatrix}
\;,
\end{equation}
namely $\Tt^E_n$ is $\Gg$-unitary (see Appendix~\ref{app-Gunitary}). The solutions \eqref{eq-transrel} can either be obtained for vectors $\phi_n$ or for operators $\phi_n$, and of particular interest is the case where the $\phi_n$ are operators of the same type as the $A_n$ and $B_n$.  A particular solution of this type is produced by
\begin{equation}
\label{eq-L2Solution}
\Phi^z_n\
\;=\;
\binom{A_{n+1}\widehat{G}^z(n+1,1)}{\widehat{G}^z(n,1)}
\;,
\qquad
\Phi^z_0\
\;=\;
\binom{\widehat{G}^z}{-\one}
\;.
\end{equation}
For $z$ not in the spectrum of $\widehat{H}$, this solution decays at $+\infty$ and it is known to be the only operator solution with this feature. All of the above applies equally well to the half-sided block Jacobi operator $\widehat{H}_{N+1}$. Its initial condition at site $N$ leading to a decaying solution is given by
$$
\widetilde{\Phi}^z_N\
\;=\;
\binom{\widetilde{G}^z_{N+1}}{-\one}
\;.
$$
As $\widehat{H}_{N+1}$ is merely a part of $\widehat{H}$ and thus has the same decaying solutions at infinity, one concludes that there is an invertible operator $C$ such that 
$$
\Phi^z_N
\;=\;
\widetilde{\Phi}^z_N\,C
\;.
$$
This identity just states that the two subspaces spanned by $\Phi^z_N$ and $\widetilde{\Phi}^z_N$ coincide, a property that can also be read off using the stereographic projection defined by:
\begin{equation}
\label{eq-StereoProj}
\Pi(\Phi)
\;=\;
(a-\imath b)(a+\imath b)^{-1}
\;,
\qquad
\Phi\;=\;\begin{pmatrix} a \\ b \end{pmatrix}
\;.
\end{equation}
One then has $\Pi(\Phi_{N}^z)=\Pi(\widetilde{\Phi}^z_N)$. As $\Phi_{N}^z=\Tt^z_{N}\cdots\Tt^z_1\Phi^z_0$, one obtains
$$
\Pi(\Tt^z_{N}\cdots\Tt^z_1\Phi^z_0)
\;=\;
(\widetilde{G}^z_{N+1}+\imath\,\one)(\widetilde{G}^z_{N+1}-\imath\,\one)^{-1}
\;=\;
(\widetilde{V}^z_{N+1})^{-1}
\;.
$$
The l.h.s. can also be expressed in terms of $\widehat{G}^z$ because the action of the matrix $\Tt^z_{N}\cdots\Tt^z_1$ is under the stereographic projection, implemented by the M\"obius action with the Cayley transform ({\it e.g.} \cite{SB})
\begin{equation}
\label{eq-MoebAct}
\Mm^z\cdot (\widehat{V}^z_1)^{-1}
\;=\;
(\widetilde{V}^z_{N+1})^{-1}
\;,
\qquad
\Mm^z
\;=\;\Cc\Tt^z_{N}\cdots\Tt^z_1\Cc^*
\;.
\end{equation}
Here the Cayley transform is
\begin{equation}
\label{eq-CayleyDef}
\Cc\;=\;2^{-\frac{1}{2}}
\begin{pmatrix}
\one & -\imath\one \\
\one & \imath\one
\end{pmatrix}
\;,
\end{equation}
and the M\"obius action $\cdot$ is defined by
\begin{equation}
\label{eq-MoebDef}
\begin{pmatrix}
a & b \\ c & d
\end{pmatrix}
\cdot
Z\;=\;
(aZ+b)(cZ+d)^{-1}
\;,
\end{equation}
provided the appearing inverse exists. For $z=E\in\RM$, this is the case because $\Mm^z$ is in the generalized Lorentz group $\mbox{U}(\Jj)$ of operators conserving the quadratic form $\Jj=\diag(\one,-\one)=\Cc\Gg\Cc^*$, see the proof of the Lemma~\ref{lem-MoebiusChern} below, which also concludes the proof of $\widetilde{V}^z_{N+1}\sim \widehat{V}^z_1$.
\hfill $\Box$

\begin{lemma}
\label{lem-MoebiusChern} 
Let $\Mm\in\mbox{U}(\Jj)\cap \Aa_{d-1}\otimes \CM^{2\times 2}$ and $\widehat{V}\in \Aa_{d-1}$ unitary. Then
$$
[\Mm\cdot \widehat{V}]_1
\;=\;
[ad^{-1}]_1
\;+\;
[\widehat{V}]_1
\;,
$$
where
$$
\Mm\;=\;
\begin{pmatrix}
a & b \\ c & d
\end{pmatrix}
\;.
$$
If $\Mm=\Cc\,\Tt^\mu\,\Cc^*$ where $\Tt^\mu$ is a transfer matrix at $\mu\in\RM$ defined as in \eqref{eq-transdef} then $[ad^{-1}]_1=0$.
\end{lemma}

\noindent {\bf Proof.} Let us recall the following facts ({\it e.g.} Lemma~2 of \cite{SB}). The identity $\Mm\Jj\Mm^*=\Jj$ implies $aa^*-bb^*=\one$ which shows that $aa^*\geq \one $ so that $a$ is invertible. Furthermore, one deduces $(a^{-1}b) (a^{-1}b)^*=\one-a^{-1}(a^{-1})^*<\one$ so that $\|a^{-1}b\|<1$. Similarly, $d$ is invertible and $\|d^{-1}c\|<1$. Now
$$
\Mm\cdot \widehat{V}
\;=\;
a\widehat{V}(\one+\widehat{V}^*a^{-1}b)(\one+d^{-1}c\widehat{V})^{-1} d^{-1}
\;,
$$
and $\one+\widehat{V}^*a^{-1}b$ as well as $\one+d^{-1}c\widehat{V}$ are homotopic to the identity, the first claim follows. Calculating $\Mm=\Cc\Tt^\mu\Cc^*$ explicitly in terms of the entries $A$ and $B=B^*$ of $\Tt^\mu$, one finds
\begin{align*}
ad^{-1}
&
\;=\;
\big((\mu-B)A^{-1}-\imath (A^{-1}+A^*)\big)\big((\mu-B)A^{-1}+\imath (A^{-1}+A^*)\big)^{-1}
\\
&
\;=\;
\big((\mu-B)-\imath (\one+A^*A)\big)\big((\mu-B)+\imath (\one+A^*A)\big)^{-1}
\;.
\end{align*}
One can deform $\mu-B$ to $0$ and $A^*A$ to $\one$, showing the last claim.
\hfill $\Box$

\subsection{Numerical procedure}
\label{sec-NumProc}

In dimension $d=2$, Theorem~\ref{theo-DimReduction} can and has been used to numerically compute Chern numbers as winding numbers \cite{ASV,AEG} when $H\cong\int^\oplus de^{\imath k}\,H(e^{\imath k})$ with a periodic one-dimensional operator $H(e^{\imath k})$. Indeed, when $H(e^{\imath k})$ is a periodic operator on $\ell^2(\ZM,\CM^L)$, then one can calculate $\widehat{V}^z(e^{\imath k})$ as the stereographic projection of the contracting dimensions of the transfer matrix over one period. This transfer matrix is a $2L\times 2L$ matrix which for real $z$ is $\Gg$-unitary. Hence its spectrum is invariant under the reflection on the unit circle and exactly half of its eigenvalues lie inside the unit disc, so that the corresponding eigenspace of these contracting eigenvalues is of dimension $L$ and thus has a well-defined stereographic projection.  Exactly the same procedure also allows to determine $\widehat{V}^\mu(e^{\imath k})$ in higher even dimension $d$, but then one has to compute a higher-dimensional winding number from $e^{\imath k}\in\TM^{d-1}\mapsto \widehat{V}^\mu(e^{\imath k})$ which in itself may then be a tough task (albeit of codimension $1$). For $d=4$, one can compute $\Ch_3(\widehat{V}^\mu)$ using the techniques of \cite{HAF} or alternatively the spectral localizer \cite{LS0}. This then provides the second Chern number $\Ch_4(P)$ of the $4$-dimensional Hamiltonian.

\section{Scattering on an insulator}
\label{sec-scat}

\subsection{Scattering set-up}
\label{sec-Q1D}

The Hamiltonian describing a semi-infinite wire coupled to a semi-infinite insulator is of the form
\begin{equation}
\label{eq-HDecomp}
\HScat\;=\;\HWireH\oplus\HInsH\,+\,\HCoup
\;.
\end{equation}
Here $\HInsH\in\widehat{\Aa}_d$ is the half-space restriction of a Hamiltonian $\HIns\in\Aa_d$ as described in Section~\ref{sec-DimRedu}. Also $\HWireH$ is an operator of the same type, albeit on a left rather then right half-space, and it will describe a conducting wire in the sense described in Section~\ref{sec-Wire} below. Hence $\HScat$ acts on the Hilbert space $\ell^2(\ZM^d,\CM^L)\cong \big(\ell^2(\NM_-)\oplus\ell^2(\NM)\big)\otimes\ell^2(\ZM^{d-1},\CM^L)$ where $\ZM=\NM_-\cup\NM$ is decomposed in $\NM_-=\{0,-1,-2,\ldots\}$ and $\NM=\{1,2,\ldots\}$. The free reference Hamiltonian (for a two Hilbert space scattering set-up) is then $\HWire\oplus\HIns$, but this will not be used below as we will essentially work in a standard quasi-one-dimensional scattering formalism as used in the solid state community. The most important point is that the focus will be on an energy $E$ which lies in the gap of $\HIns$ and for which the wire is perfectly conducting, see Section~\ref{sec-Wire} below for the latter. This implies that the scattering matrix $S^E$ at this energy only consists of a unitary reflection matrix as the transmission matrix vanishes. To further simplify, we suppose that the system can be diagonalized by a Fourier transform along the boundary, as in \eqref{eq-HHastDiag}:
$$
\HScat\;\cong\;
\int^\oplus_{\TM^{d-1}} dk\;
\HScat(e^{\imath k})\;=\;
\int^\oplus_{\TM^{d-1}} dk\;
\big(\HWireH\oplus\HInsH(e^{\imath k})\,+\,\HCoup\big)
\;.
$$
In particular, $\HWireH$ and $\HCoup$ are supposed to be independent of $k$. This means that there are infinitely many independent wires attached to the half-sided insulator with local and identical coupling terms. Each wire is then described by a periodic block Jacobi matrix of the form
\begin{equation} 
\label{eq-HWire}
\HWire\;=\;A\,S\,+B\,+\,A^*\,S^*
\;,
\end{equation}
where $A$ and $B=B^*$ are $L\times L$ matrices with $A$ invertible, and $S$ is the left-shift on $\ell^2(\ZM)$. Under further hypothesis on $A$ and $B$, this operator will be analyzed in Section~\ref{sec-Wire}. For sake of simplicity, the coupling of the wire to the insulator is expressed in terms of the same invertible matrix $A$:
$$
\HCoup\;=\;\big(\pi_0^*\,A\,\pi_1\,+\,\pi_1^*\,A^*\,\pi_0\big)
\;,
$$
where $\pi_j$ is the partial isometries onto the fiber $\CM^L$ over the site $j$. 
Summing up, a matrix representation of the quasi-one-dimensional scattering Hamiltonian at $e^{\imath k}\in\TM^{d-1}$ is
\begin{equation}
\label{eq-Hrep}
\HScat(e^{\imath k})
\;=\;
\begin{pmatrix}
\ddots & \ddots & & & \\
\ddots & B & A & & \\
& A^* & B_1(k) & A_2(k) & \\
& & A_2(k)^* & B_2(k) & \ddots \\
& &  & \ddots & \ddots
\end{pmatrix}
\;.
\end{equation}
The picture in Section~\ref{sec-ScatStates} below also illustrates this Hamiltonian. The next sections analyze this parametrized family of quasi-one-dimensional scattering Hamiltonians. This allows to prove Theorem~\ref{theo-ScatInv}.

\subsection{Description of the wire}
\label{sec-Wire}

The Hamiltonian $\HWire$ is diagonalized by the Fourier transform 
$$
\Ff:\ell^2(\ZM,\CM^L)\to L^2\left(\TM^1,\frac{dp}{2\pi}\right)\otimes \CM^L\;,
\qquad
(\Ff\psi)(p)\;=\;\sum_{n\in\ZM}\psi_{n} e^{\imath pn}\;,
$$
namely
$$
\Ff \HWire\Ff^*\;=\;\int_{\TM^1}\frac{d p}{2\pi}\,\HWire(e^{\imath p})\;,
\qquad
\HWire(e^{\imath p})\;=\;A\,e^{-\imath p}\,+B\,+\,A^*\,e^{\imath p}
\;.
$$
Note that the matrix $\HWire(e^{\imath p})$ is self-adjoint. By analytic perturbation theory, $\HWire(e^{\imath p})$ has $L$ eigenvalues $E_l(p)$, $l=1,\ldots,L$, which at level crossings can be chosen to be analytic. If $\phi_l(p)$ are the corresponding eigenvectors, one has
\begin{equation}
\label{eq-Bloch}
\HWire(e^{\imath p})\,\phi_l(p)
\;=\;
E_l(p)\,\phi_l(p)\;,
\qquad
l=1,\ldots,L\;,
\end{equation}
As $p$ runs through $[-\pi,\pi)$ each eigenvalue leads to an energy band of $\HWire$. The spectrum of $\HWire$ is absolutely continuous if none of the eigenvalues is constant in $p$. If such a constant energy occurs, then one also speaks of a flat band. It leads to a Dirac peak in the density of states. 
It can be shown that there are no flat bands if $A$ is invertible. The latter is a standing assumption. Now let $\Tt^E=\Tt^E_0=\Tt^E_n$ for $n\leq 0$ be the transfer matrix at energy $E\in\RM$ given by \eqref{eq-transdef}. There is a simple link between the eigenvalues of the transfer matrix on the unit circle and the Bloch solutions \eqref{eq-Bloch}.

\begin{proposi}
\label{prop-eigenvectors}
One has the following equivalence:
$$
\HWire(e^{\imath p})\,\phi\;=\;E\,\phi
\qquad
\Longleftrightarrow
\qquad
\Tt^E\binom{e^{\imath p}A\,\phi}{\phi}\;=\;e^{\imath p}\binom{e^{\imath p}A\,\phi}{\phi}
\;.
$$
\end{proposi}

\noindent {\bf Proof.} The lower equation of the second equality is trivially verified. The upper one is precisely the Schr\"odinger equation on the left multiplied by $e^{\imath p}$.
\hfill $\Box$

\vspace{.2cm}

The transfer matrix $\Tt^E$ is a $\Gg$-unitary and hence satisfies all the properties listed in Appendix~\ref{app-Gunitary}. In particular, it is possible to associated Krein signatures to every eigenvalue on the unit circle, see Definition~\ref{def-signatures} in Appendix~\ref{app-Gunitary}. The following will be supposed throughout.

\vspace{.2cm}

\hypertarget{hyp}
\noindent {\bf Hypothesis:} {\it The wire is perfectly conducting, namely its transfer matrix has only eigenvalues on the unit circle which are Krein definite in the sense of {\rm Definition~\ref{def-signatures}}.}

\vspace{.2cm}

Krein collision theory of unit eigenvalues states that the signature is conserved when eigenvalues leave the unit circle (see \cite{Kre,GLR,SB3}). Having only definite eigenvalues is therefore a stability property which is part of the assumption on the wire.  The hypothesis also implies that the transfer matrix has no parabolic part corresponding to non-trivial Jordan blocks for eigenvalues on the unit circle (see Corollary~\ref{coro-signaturesJordan} in Appendix~\ref{app-Gunitary}). As there is no hyperbolic part corresponding to eigenvalues off the unit circle neither, the transfer matrix is fully elliptic. In physical terms: the wire has only open or conducting channels so that there are no closed or evanescent channels. This allows to diagonalize the transfer matrix  just as in \eqref{eq-PsiDef1} by $2L\times L$ matrices $\Psi^E_\pm$ of rank $L$ consisting of eigenvectors for eigenvalues of positive/negative signature. Using the Cayley transform \eqref{eq-CayleyDef} as in \eqref{eq-PsiDef2}, the basis change $\Nn^E=(\Psi^E_+,\Psi^E_-)\Cc $ satisfies
\begin{equation}
\label{eq-PsiDef2bis}
(\Nn^E)^*\Gg\Nn^E
\;=\;
\Gg
\;,
\qquad
(\Nn^E)^{-1}\Tt^E\Nn^E
\;=\;
\tfrac{1}{2}
\begin{pmatrix}
\Lambda^E_++\Lambda^E_- & -\imath(\Lambda^E_+-\Lambda^E_-) \\ \imath(\Lambda^E_+-\Lambda^E_-) & \Lambda^E_++\Lambda^E_-
\end{pmatrix}
\;,
\end{equation}
where $\Lambda^E_\pm$ are diagonal $L\times L$ matrices. Thus $\Nn^E$ is a $\Gg$-unitary diagonalizing $\Tt^E$ to a direct sum of real $2\times 2$ rotation matrices multiplied by a phase factor. The entries of $\Nn^E$ are also denoted as follows
$$
\Nn^E
\;=\;
(\Psi^E_\lor,\Psi^E_\wedge)
\;=\;
\;=\;
2^{-\frac{1}{2}}\big(\Psi^E_++\Psi^E_-, -\imath(\Psi^E_+-\Psi^E_-)\big)
\;.
$$
Both $\Psi^E_\lor$ and $\Psi^E_\wedge$ span $\Gg$-Lagrangian subspaces, and they are $\Gg$-orthogonal to each other. The (Krein) signature of unit eigenvalues of $\Tt^E$ also determines the sign of the group velocity. For this purpose, one has to analyze the dependence of $\Tt^E$ and its eigenvalues on $E$. The following result also shows that eigenvalues with positive/negative signature leave the unit circle to the inside/outside as a positive imaginary part is added to the energy, namely $\zeta$ contains an imaginary part.

\begin{proposi}
\label{prop-eigenphasederivative}
Let $\lambda^E$ be a unit eigenvalue of the transfer matrix $\Tt^E$ with eigenspace $\Ee^E_{\lambda^E}$. Further let $I\subset\RM$ be an open interval such that $E\in I\mapsto\dim(\Ee^E_{\lambda^E})=1$ is constant. Then
\begin{equation}
\label{eq-eigenphasederivative}
\lambda^{E+\zeta}
\;=\;
\lambda^E\,
\exp\left(
\imath\;
\frac{\frac{1}{\imath}\,(v^E)^*(\Tt^E)^*\Gg(\partial_E\Tt^E)v^E}{(v^E)^*\Gg v^E}\,\zeta
\,+\,
\Oo(\zeta^2)
\right)
\;,
\end{equation}
where $v^E$ is the eigenvector of $\Tt^E$ to the eigenvalue $\lambda^E$. Moreover,  
$$
\frac{1}{\imath}(v^E)^*(\Tt^E)^*\Gg(\partial_E\Tt^E)v^E
\;>\;0\;.
$$
Hence the phase $\theta^E=-\imath\log(\lambda^E)$ satisfies
$$
\mbox{\rm sign}(\partial_E\theta_E)
\;=\;
\nu_+(\lambda^E)-\nu_-(\lambda^E)
\;.
$$
\end{proposi}

\noindent {\bf Proof.} By analytic perturbation theory, $\lambda^E$ and $v^E$ are analytic in $E$ (with proper choices of branches, this holds even at level crossings,  see Section III.1.5 of \cite{YS}).  Deriving the eigenvalue equation $\Tt^Ev^E=\lambda^Ev^E$, one finds
$$
(\partial_E\Tt^E)v^E\,+\,\Tt^E(\partial_Ev^E)
\;=\;
(\partial_E\lambda^E)v^E\,+\,\lambda^E(\partial_Ev^E)
\;.
$$
Now multiplying this equation from the left by $(v^E)^*(\Tt^E)^*\Gg$. As $(v^E)^*(\Tt^E)^*\Gg\Tt^E=\overline{\lambda}^E(v^E)^*\Gg\Tt^E$, it follows that
$$
(v^E)^*\Gg(\partial_E\Tt^E)v^E
\;=\;
(\partial_E\lambda^E)\;(v^E)^*\Gg v^E
\;.
$$
This allows to calculate $\partial_E\lambda^E$ and therefore also \eqref{eq-eigenphasederivative}. Furthermore,
$$
\frac{1}{\imath}
\;(\Tt^E)^*\Gg(\partial_E\Tt^E)
\;=\;
\begin{pmatrix}
(A^{-1})^*(E-B) & (A^{-1})^* \\
- A & 0 
\end{pmatrix}
\begin{pmatrix}
0 & -\one \\
\one & 0 
\end{pmatrix}
\begin{pmatrix}
A^{-1} & 0 \\
0 & 0 
\end{pmatrix}
\;=\;
\begin{pmatrix}
(A^{-1})^*A^{-1} & 0 \\
 0 & 0
\end{pmatrix}
\;.
$$
As the eigenvectors of $\Tt^E$ never have a vanishing upper component, this shows the claimed positivity and therefore concludes the proof.  From this the last fact follows.
\hfill $\Box$

\subsection{Scattering states and reflection matrix}
\label{sec-ScatStates}

Scattering states of $\HScat$ have to be constructed by matching $\Phi^E$ with the bounded solutions $\Psi^E_+$ and $\Psi^E_-$ in the wire as given in Section~\ref{sec-Wire}. This matching can be done in a purely geometric way that leads to a definition of the reflection matrix, by applying Proposition~\ref{prop-angles} in the Appendix~\ref{app-Gunitary} to $\Phi^E$,  $\Psi^E_+$ and $\Psi^E_-$. In this section, it is verified that this procedure leads to the reflection matrix of the scattering set-up described in Section~\ref{sec-Q1D}. Indeed, there is only a reflection of incoming states from the wire to outgoing states into the wire, namely there are no transmitted states. The incoming states are those having a positive group velocity and, due to Proposition~\ref{prop-eigenphasederivative}, are thus corresponding to those initial conditions generated from eigenvectors $\Psi^E_+$ of the transfer matrix $\Tt^E$ with positive Krein signature (at fixed energy $E$). Similarly, the outgoing states use initial conditions with negative Krein signature. Hence incoming/outgoing plane waves $(\psi_{\pm,n})_{n\leq 0}$ at energy $E$  are given in terms of vectors $w_\pm\in\CM^L$ by 
$$
\psi_{\pm,n}
\;=\;
\binom{0}{\one}^*(\Tt^E)^{n}\Psi^E_\pm w_\pm\;\in\;\CM^L
\;.
$$
Up to now, the plane wave states are generalized eigenstates of the Schr\"odinger equation at energy $E$ in the wire only, and not yet of the Hamiltonian $\HScat$ of the wire coupled to the insulator. To get such solutions $\phi$ of $\HScat\phi=E\phi$, one has to take linear combinations of $\psi_+$ and $\psi_-$ in such a manner that they match the $\ell^2$-solutions in the insulator. Due to the nearest neighbor hopping, this matching can be studied at the sites $0$ and $1$ via
\begin{equation}
\label{eq-ScatMatch}
\Psi^E_+w_+\,+\,\Psi^E_-w_-\;=\;\Phi^E\,w_{\mbox{\rm\tiny Ins}}
\;,
\end{equation}
for some suitable $w_{\mbox{\rm\tiny Ins}}\in\CM^L$. Here $\Phi^E$ is the initial condition of all decaying solutions of $ \HInsH\phi=E\phi$ in the insulator. Starting from the initial condition $\Phi^E$, it is obtained in the insulator via the transfer matrices as in \eqref{eq-transrel}. The matching \eqref{eq-ScatMatch} is illustrated in the following picture.

\vspace{-1cm}

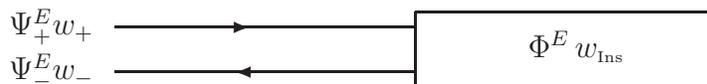
\setlength{\unitlength}{1mm} 
\begin{picture}(140,40)(-90,-5)

\thicklines
\put(0,10){\line(1,0){40}}
\put(0,20){\line(1,0){40}}
\put(0,10){\line(0,1){10}}
\put(0,12){\line(-1,0){40}}
\put(0,18){\line(-1,0){40}}
\put(-24,12){\vector(-1,0){0}}
\put(-22,18){\vector(1,0){0}}


\thinlines
\put(-54,17){$ \Psi^E_+ w_+ $}
\put(-54,11){$ \Psi^E_- w_- $}
\put(15,14){$ \Phi^E\,w_{\mbox{\rm\tiny Ins}} $}
\end{picture}

\vspace{-1cm}

\noindent All states constructed from this matching are called scattering states. As the connection between these incoming and outgoing states is linear, one then defines the reflection matrix $R^E$ at energy $E$ by:
\begin{equation}
\label{eq-ScatMatch2}
w_-\;=\;R^E\,w_+
\;.
\end{equation}
As $\Phi^E$ spans an $L$-dimensional subspace, equation \eqref{eq-ScatMatch} shows that there is at most an $L$-dimensional space of such solutions.  In fact, it is precisely of dimension $L$ (for an elliptic wire). Then there is a purely geometric way of extracting a unitary matrix. In order to stress this, Proposition~\ref{prop-angles} in Appendix~\ref{app-Gunitary} restates the following result in that context.

\begin{proposi}
\label{prop-RelationRU} 
Suppose that the wire satisfies the above \hyperlink{hyp}{\rm Hypothesis}. Then the space of scattering states is of dimension $L$ and spanned by $\Psi^E_++\Psi^E_-R^E$ where $R^E$ is a unitary $L\times L$ matrix called the reflection matrix at energy $E$. It is given by a M\"obius transformation of the stereographic projection $\Pi(\Phi^E)$: 
$$
(R^E)^{-1}
\;=\;
\big(\Cc(\Nn^E)^{-1}\Cc^*\big)\cdot \Pi(\Phi^E)
\;.
$$
In terms of $\widehat{V}^E=(\widehat{G}^E-\imath\one)(\widehat{G}^E+\imath\one)^{-1}$, one has 
\begin{equation}
\label{eq-RFormula}
R^E
\;=\;
\big(\overline{\Cc}(\Nn^E)^{-1}\Cc^T\big)\cdot\widehat{V}^E
\;.
\end{equation}
\end{proposi}

\noindent {\bf Proof.} The equation \eqref{eq-ScatMatch} can be written in a matrix form as
$$
(\Psi^E_+,\Psi^E_-)\begin{pmatrix} W_+ \\ W_- \end{pmatrix}
\;=\;
\Phi^E
\;,
$$
with $L\times L$ matrices $W_\pm$. Next recall that $\Nn^E=(\Psi^E_+,\Psi^E_-)\Cc $ is a $\Gg$-unitary, so that
$$
(\Psi^E_+,\Psi^E_-)^*\Gg(\Psi^E_+,\Psi^E_-)
\;=\;
\Jj
\;,
$$
where $\Jj=\diag(\one,-\one)$ as above. Hence $(\Psi^E_+,\Psi^E_-)^{-1}=\Jj(\Psi^E_+,\Psi^E_-)^*\Gg$ so that
$$
\begin{pmatrix} W_+ \\ W_- \end{pmatrix}
\;=\;
\Jj(\Psi^E_+,\Psi^E_-)^*\Gg\Phi^E
\;.
$$
Moreover, $(\Psi^E_+,\Psi^E_-)^{-1}$ maps the $\Gg$-Lagrangian subspace $\Phi^E$ to a $\Jj$-Lagrangian subspace. This implies that $W_+^*W_+=W_-^*W_-$ and that this operator is invertible. Consequently $W_-W_+^{-1}$ is unitary. Writing
$$
\begin{pmatrix} W_+ \\ W_- \end{pmatrix}
\;=\;
\begin{pmatrix} \one \\ W_-W_+^{-1} \end{pmatrix}W_+
\;=\;
\begin{pmatrix} \one \\ R^E \end{pmatrix}W_+
\;,
$$
this shows the first claim.  On the other hand, using the stereographic projection \eqref{eq-StereoProj} and the fact that matrix multiplication with $\Gg$-unitary  before stereographic projections becomes M\"obius action with the Cayley transform (just as in \eqref{eq-MoebAct})
\begin{align*}
\Pi(\Phi^E)
& \;=\;
\Pi\big(\Psi^E_++\Psi^E_-R^E\big)
\;=\;
\Pi\Big(\Nn^E\Cc^*\binom{\one}{R^E}\Big)
\;=\;
\big(\Cc\Nn^E\Cc^*\big)\cdot (R^E)^{-1}
\;,
\end{align*}
which after inversion of the M\"obius transformation shows the second claim. The last formula follows since the decaying solutions in the insulator are given by \eqref{eq-L2Solution} so that $\Pi(\Phi^E)=(\widehat{V}^E)^{-1}$. Hence \eqref{eq-Moeb} implies
$$
(R^E)^{-1}
\;=\;
\big(\Cc(\Nn^E)^{-1}\Cc^*\big)\cdot(\widehat{V}^E)^{-1}
\;.
$$
Calculating the inverse completes the proof.
\hfill $\Box$

\vspace{.2cm}

For the following, it is important to extend the reflection matrix $R^E$ to complex energies $z=E+\imath\delta$ with small imaginary part by using \eqref{eq-RFormula}. The existence of an analytic continuation of $\widehat{V}^E$ is guaranteed, under the stated conditions, by Proposition~\ref{prop-CayleyInvertible}. Furthermore, recall that $\Nn^E$ is merely obtained from the spectral analysis of the transfer matrix $\Tt^E$ of the wire and therefore it also has an analytic extension off the real axis. Hence \eqref{eq-RFormula} becomes
\begin{equation}
\label{eq-RFormula2}
R^z
\;=\;
\big(\overline{\Cc}(\Nn^z)^{-1}\Cc^T\big)\cdot\widehat{V}^z
\;.
\end{equation}
Let us stress that $R^z\in\Ee_d$ and that it is invertible and differentiable as long as Proposition~\ref{prop-CayleyInvertible} holds. Finally let us note that it is possible to rewrite \eqref{eq-RFormula2} in terms of the Green matrices of the half-sided wire and the half-sided insulator. For simplicity, let us give the outcome in the case of $A=\one$ and $B=0$:
$$
R^z\;=\;
\big(\widehat{G}^z_{\mbox{\rm\tiny Wire}}-\widehat{G}^z_{\mbox{\rm\tiny Ins}}\big)
\big(\widehat{G}^z_{\mbox{\rm\tiny Wire}}+\widehat{G}^z_{\mbox{\rm\tiny Ins}}\big)^{-1}
\;.
$$
Similar expressions have in one-dimensional context been used for a detailed spectral analysis of the underlying operators \cite{Rem}.

\subsection{Odd Chern number of the reflection matrix}
\label{sec-ReflMatrix}

\noindent {\bf Proof} of Theorem~\ref{theo-ScatInv}. Let us first consider the case $\lambda=0$ so that $\widehat{H}$ is given by \eqref{eq-HHastDiag}. Due to \eqref{eq-RFormula}, one can apply Lemma~\ref{lem-MoebiusChern} with $\Mm=\overline{\Cc}(\Nn^E)^{-1}\Cc^T$ to deduce that
$$
[R^E]_1
\;=\;
[ad^{-1}]_1\,+\,[\widehat{V}^E]_1
\;,
$$
where $a$ and $d$ are the diagonal entries of $\Mm$. As $\Nn^E$ and thus also $a$ and $d$ are independent of $k$, it follows that
$$
\Ch_{d-1}(R^E)
\;=\;\Ch_{d-1}(\widehat{V}^E)
\;.
$$
Now one can analytically extend both sides as in  \eqref{eq-RFormula2} by using Proposition~\ref{prop-CayleyInvertible}. Combined with Theorem~\ref{theo-DimReduction} this completes the proof of Theorem~\ref{theo-ScatInv}.
\hfill $\Box$

\appendix

\section{Cayley transformation of operators}
\label{app-Cayley}

Let $\Hh$ be a separable Hilbert space and $\BM(\Hh)$ the bounded operators on $\Hh$. Let us define the upper half plane and unit disc of operators on $\Hh$ by
\begin{align*}
& \UM(\Hh)
\;=\;
\left\{
Z=X+\imath Y\,:\,X=X^*\in\BM(\Hh)\;,\;\;0<Y\in\BM(\Hh)
\right\}
\;,
\\
& \DM(\Hh)
\;=\;
\left\{
Z\in\BM(\Hh)\,:\,Z^*Z<\one
\right\}
\;.
\end{align*}
Recall that an operator $Z$ with positive imaginary part $Y=\frac{1}{2\imath}(Z-Z^*)$ is invertible.

\begin{proposi}
\label{prop-Cayley} 
The Cayley transform $\Cc\cdot$ defined as the M\"obius transformation with the matrix $\Cc$ given in \eqref{eq-CayleyDef} by
$$
\Cc\cdot Z
\;=\;
(Z-\imath)(Z+\imath)^{-1}
$$
is a bijection from the upper half-plane $\UM(\Hh)$ to the unit disc $\DM(\Hh)$.
\end{proposi}

\noindent {\bf Proof.} Clearly
$$
\Cc\cdot Z
\;=\;
\one\,-\,2\imath(Z+\imath)^{-1}
\;.
$$
Hence $|\Cc\cdot Z|^2=(\Cc\cdot Z)^*\Cc\cdot Z\geq 0$ can be multiplied out to
\begin{align*}
|\Cc\cdot Z|^2
&
\;=\;
\one\,+\,2\imath(Z^*-\imath)^{-1}
\,-\,2\imath(Z+\imath)^{-1}
\,+\,4(Z^*-\imath)^{-1}(Z+\imath)^{-1}
\\
&
\;=\;
\one\,-\,4\big((Z+\imath)^{-1}\big)^*Y(Z+\imath)^{-1}
\;.
\end{align*}
As $Y>0$, one indeed concludes $|\Cc\cdot Z|^2<\one$. The inverse map is given by the M\"obius transformation with $\Cc^*$.
\hfill $\Box$

\section{$\Gg$-unitaries and some of their properties}
\label{app-Gunitary}

The transfer matrices are $\Gg$-unitary, namely they satisfy \eqref{eq-GUnitarity}. Here $\Gg$ is a non-degenerate indefinite sesquilinear form with inertia $(L,L)$. This appendix collects a few facts about the spectral properties of a $\Gg$-unitary $\Tt$ coming from finite dimensional Krein space theory, see \cite{Kre,YS,GLR} and also \cite{SB3}. 

\vspace{.2cm}

\noindent (1) If $\lambda$ is an eigenvalue of $ \Tt $ then also $(\overline{\lambda})^{-1}$ is an eigenvalue. 

\vspace{.1cm}

\noindent (2) Two generalized eigenspaces $\Ee_\lambda$ and $\Ee_\mu$ of $\lambda$ and $\mu$ are $\Gg$-orthogonal if $\lambda^{-1}\not =\overline{\mu}$, meaning that $v^*\Gg w=0$ for all $v\in \Ee_\lambda$ and $w\in\Ee_\mu$.


\begin{defini}
\label{def-signatures}
Let $\Tt$ be a $\Gg$-unitary and $\lambda$ be an eigenvalue of modulus $1$. Its (Krein) signature $(\nu_+(\lambda),\nu_-(\lambda))$ is defined as the signature of the sesquilinear form $\Gg$ restricted to the generalized eigenspace $\Ee_\lambda$. Eigenvalues for which either $\nu_+(\lambda)=0$ or $\nu_-(\lambda)=0$ are called definite or of definite signature. An eigenvalue which is not definite is called indefinite or of mixed signature.
\end{defini}

The signature is the central concept of Krein collision theory of unit eigenvalues which essentially states that the signature in the above sense has to be conserved when eigenvalues leave the unit circle (see \cite{Kre,GLR,SB3}). The following result shows that for eigenvalues on the unit circle the definiteness can be checked by only looking at eigenvectors (and hence not at the generalized eigenvectors). 

\begin{proposi}
\label{prop-definitenesscheck}
Let $\lambda$ be an eigenvalue of a $\Gg$-unitary $\Tt$ with $|\lambda|=1$. Then, for both $\sigma=\pm$,
$$
\nu_\sigma(\lambda)\;=\;0
\qquad
\Longleftrightarrow
\qquad
(-\sigma)\;v^*\Gg v \;>\; 0
\;\;\mbox{ for all eigenvectors }v\mbox{ of }\lambda\;.
$$
\end{proposi}

\noindent {\bf Proof.} The implication $\Longrightarrow$ is clear. For the converse, we show that the condition on the r.h.s. actually implies that $\Ee_\lambda$ only consists of eigenvectors so that again the definiteness follows. Hence let us suppose that there is a non-trivial Jordan block, namely that there are vectors $v$ and $w$ such that $\Tt v=\lambda v$ and $\Tt w=\lambda w+v$. Then
$$
v^*\Gg w
\;=\;
v^*\Tt^*\Gg\Tt w
\;=\;
(\lambda v)^*\Gg(\lambda w+v)
\;=\;
|\lambda|^2 v^*\Gg w+\overline{\lambda} v^*\Gg v
\;.
$$
Hence
$$
\overline{\lambda} v^*\Gg v
\;=\;
(1-|\lambda|^2) v^*\Gg w
\;=\;
0
\;.
$$
But this shows $v^*\Gg v=0$, which is a contradiction to the hypothesis.
\hfill $\Box$

\vspace{.2cm}

The previous proof actually also shows the following result.

\begin{coro}
\label{coro-signaturesJordan}
Let $\Tt$ be a $\Gg$-unitary and $\lambda$ be an eigenvalue of modulus $1$. 

\vspace{.1cm}

\noindent {\rm (i)} If there is a non-diagonal Jordan block for $\lambda$, then $\lambda$ is indefinite and there exists an 

eigenvector $v$ such that $v^*\Gg v=0$.

\vspace{.1cm}

\noindent {\rm (ii)} If $\lambda$ is definite, then all Jordan blocks are diagonal.
\end{coro}

Next let us consider a $\Gg$-unitary $\Tt$ which is purely elliptic, namely it has only spectrum on the unit circle with trivial Jordan blocks. We will bring this matrix in its normal form. There exist $2L\times L$ matrices $\Psi_\pm$ of rank $L$ consisting of eigenvectors for eigenvalues of positive/negative signature such that
\begin{equation}
\label{eq-PsiDef1}
(\Psi_+,\Psi_-)^*\Gg(\Psi_+,\Psi_-)
\;=\;
\Jj
\;,
\qquad
(\Psi_+,\Psi_-)^{-1}\Tt(\Psi_+,\Psi_-)
\;=\;
\begin{pmatrix}
\Lambda_+ & 0 \\ 0 & \Lambda_-
\end{pmatrix}
\;,
\end{equation}
where $\Jj=\diag(\one,-\one)$ and $\Lambda_\pm$ are diagonal $L\times L$ matrices. Using the Cayley transform \eqref{eq-CayleyDef} and the basis change $\Nn=(\Psi_+,\Psi_-)\Cc $, one can then check
\begin{equation}
\label{eq-PsiDef2}
\Nn^*\Gg\Nn
\;=\;
\Gg
\;,
\qquad
\Nn^{-1}\Tt\Nn
\;=\;
\tfrac{1}{2}
\begin{pmatrix}
\Lambda_++\Lambda_- & -\imath(\Lambda_+-\Lambda_-) \\ \imath(\Lambda_+-\Lambda_-) & \Lambda_++\Lambda_-
\end{pmatrix}
\;,
\end{equation}
namely $\Nn$ is $\Gg$-unitary diagonalizing $\Tt$ to a direct sum of real $2\times 2$ rotation matrices multiplied by a phase factor. The entries of $\Nn$ are also denoted as follows
$$
\Nn
\;=\;
(\Psi_\lor,\Psi_\wedge)
\;=\;
\;=\;
2^{-\frac{1}{2}}\big(\Psi_++\Psi_-, -\imath(\Psi_+-\Psi_-)\big)
\;.
$$
Both $\Psi_\lor$ and $\Psi_\wedge$ span $\Gg$-Lagrangian subspaces, and they are $\Gg$-orthogonal to each other.

\vspace{.2cm}

The following geometric result show that a unitary $R$ specifies the position of a $\Gg$-Lagrangian plane w.r.t. a coordinate system given by a $\Gg$-unitary $\Nn=(\Psi_\lor,\Psi_\wedge)$. When applied to the scattering set-up, this unitary is the reflection matrix.

\begin{proposi}
\label{prop-angles} 
Let ${\Phi}$ span a $\Gg$-Lagrangian subspace and let $\Nn=(\Psi_\lor,\Psi_\wedge)$ be a $\Gg$-unitary. Define $L\times L$ matrices $N$ and $M$ by 
\begin{equation}
\label{eq-coordinaterep2}
\Phi
\;=\;
\Psi_\lor\,N
\;+\;
\Psi_\wedge\,M
\;.
\end{equation}
Then the matrix
\begin{equation}
\label{eq-coordinateUdef2}
R
\;=\;
(N-\imath M)(N+\imath M)^{-1}
\;,
\end{equation}
is unitary. Furthermore $1$ is eigenvalue of $R$ if and only if $\Ff=\Ran(\Phi)$ and $\Ee=\Ran(\Psi_\lor)$ have a non-trivial intersection. More precisely,
$$
\dim\big(\Ker(R-\one)\big)
\;=\;
\dim(\Ff\cap\Ee)
\;.
$$
Using the stererographic projection $\Pi$ defined in \eqref{eq-StereoProj}, one has 
\begin{equation}
\label{eq-Moeb}
\Pi(\Phi)
\;=\;
(\Cc\Nn\Cc^*)\cdot R
\;,
\end{equation}
where $\cdot$ denotes the matrix M\"obius transformation as defined in \eqref{eq-MoebDef}.
\end{proposi}

\noindent {\bf Proof.}  First let note that $N$ and $M$ are indeed well-defined. Actually, the equation can also be rewritten as 
$$
\begin{pmatrix}
N \\ M
\end{pmatrix}
\;=\;
(\Psi_\lor,\Psi_\wedge)^{-1}
\;\Phi
\;,
$$
because $\Nn=(\Psi_\lor,\Psi_\wedge)$ is invertible. As $\Nn$ is $\Gg$-unitary, also $\binom{N}{M}$ spans a $\Gg$-Lagrangian subspace and its stereographic projection is nothing but ${R}$ which is hence unitary \cite{SB3}. Let us check that directly once again. Indeed, one has $N^*M=M^*N$ from the $\Gg$-Lagrangian property. From this follows that
$$
(N\pm \imath M)^*(N\pm\imath M)
\;=\;
N^*N+M^*M\,\pm\,\imath (N^*M-M^*N)\;=\;N^*N+M^*M
\;,
$$
so that $N\pm \imath M$ are both invertible and also
$$
(N+\imath M)^*(N+\imath M)
\;=\;
(N-\imath M)^*(N-\imath M)
\;,
$$
so that
$$
(N-\imath M)(N+\imath M)^{-1}
\;=\;
((N-\imath M)^*)^{-1}(N+\imath M)^*
\;=\;
\big(\big((N-\imath M)(N+\imath M)^{-1}\big)^*\big)^{-1}
\;,
$$
that is ${R}$ is indeed unitary.  Finally
\begin{align*}
\Ker({R}-\one)
&
\;=\;
\Ker\big((N-\imath M)(N+\imath M)^{-1}-\one\big)
\\
&
\;=\;
(N+\imath M)\;\Ker\big((N-\imath M)-(N+\imath M)\big)
\\
&
\;=\;
(N+\imath M)\;\Ker(M)
\;.
\end{align*}
But a vector $w$ in the kernel of $M$ produces a vector  $\Phi w=\Psi_\lor Nw$ in the intersection of $\Ff$ and $\Ee$. This implies the claim. The last one follows from \cite{SB2}.
\hfill $\Box$

\vspace{.3cm}

\noindent {\bf Acknowledgements:}  We thank an anonymous referee for filling a gap in one of the proofs. This research was partly supported by DFG  grant SCHU 1358/6-1.


\end{document}